\documentclass[twocolumn, aps, prl]{revtex4}
\usepackage{graphicx}
\usepackage{amssymb}
\usepackage{amsmath}

\newcommand{\GVec}[1]{\mbox{\boldmath$#1$}}

\def\Vec#1{{\bf #1}}
\def\GVec#1{\mbox{\boldmath $#1$}}

\def\H{{\mathcal H}}
\def\vare{\varepsilon}

\begin{document}

\title{Magnetic field screening and mirroring in graphene}
\author{Mikito Koshino, Yasunori Arimura, and Tsuneya Ando}
\affiliation{
Department of Physics, Tokyo Institute of Technology,
Tokyo 152-8551, Japan}
\date{\today}

\begin{abstract}
The orbital magnetism in spatially varying magnetic fields is studied in monolayer graphene within the effective mass approximation.
We find that, unlike the conventional two-dimensional electron system,
 graphene with small Fermi wave number $k_F$ works as a magnetic shield
 where the field produced by a magnetic object placed above graphene
is always screened by a constant factor on the other side of graphene.
The object is repelled by a diamagnetic force from the graphene, as if there exists its mirror image with a reduced amplitude on the other side of graphene.
The magnitude of the force is 
much greater than that of conventional two-dimensional system.
The effect disappears with the increase of $k_F$.
\end{abstract}

\maketitle


Graphene, an atomic sheet of graphite, has a peculiar
electronic structure analog to a relativistic particle,
and its unique properties have been
of great interest.
In graphene the conduction and valence bands stick together with a linear dispersion.
The low-energy physics is successfully described by the effective-mass 
Hamiltonian analogous to the massless Dirac Fermion,
\cite{Slonczewski_Weiss_1958a,DiVincenzo_Mele_1984a,Semenoff_1984a,Ando_2005a}
and its unique transport properties 
were studied.
\cite{Ando_2005a,Shon_and_Ando_1998a,Zheng_and_Ando_2002a,Gusy05,Pere06b}
Since its experimental fabrication,
\cite{Novo04,Novo05,Zhan05-2}
the graphene and related materials have attracted
much attention and
have been extensively investigated in experiments and theories.

The electronic property of graphene 
in a magnetic field was first investigated in theories
as a simple model of the bulk graphite. \cite{McCl56}
There it was shown that, in a uniform magnetic field, 
graphene exhibits a huge diamagnetic susceptibility
due to the orbital motion of electrons,
which is quite different from 
the conventional Landau diamagnetism. 
The orbital magnetism was also studied for
related materials such as
the bulk graphite, \cite{McCl60,Shar}
graphite intercalation compounds, \cite{Safr,Blin,Sait}
carbon nanotube, \cite{Ajik93,Yama08}
disordered graphene,\cite{Gusy04,Fuku07,Koshino_Ando_2007a}
few-layered graphenes, \cite{Koshino_Ando_2007b,Naka08,Cast09}
and nodal Fermions \cite{Ghos07}.
Quite recently, the graphene in a spatially modulated magnetic field
was studied in the context of 
the electron confinement using magnetic barrier.
\cite{Pere06,DeMa07,Rame07}
The diamagnetic susceptibility was experimentally 
observed for quasi-two-dimensional graphite,
which is a random stack of graphene sheets. \cite{Koto86}

In this paper, we study the orbital diamagnetism
in non-uniform magnetic fields in monolayer graphene.
Using the effective mass approximation and the perturbation theory,
we calculate the electric current
induced by an external magnetic field with wavenumber $q$,
to obtain susceptibility $\chi(q)$
for general Fermi energies.
We apply the result to arbitrary geometries
where a certain magnetic object is located above graphene,
and estimate the response current induced on graphene,
as well as the diamagnetic repulsive force which works between 
graphene and the magnetic object.
We find that graphene has a peculiar property of magnetic mirroring,
where the counter field induced by the response current 
mimics a mirror image of the original object.


We start with the general formulation of the electric response to
the spatially-varying magnetic field in a two-dimensional (2D) system.
We assume a uniform 2D system on the $xy$ plane,
and apply a magnetic field perpendicular to the layer
$B(\Vec{r}) = [\nabla \times \Vec{A}(\Vec{r})]_z$
with vector potential $\Vec{A}(\Vec{r})$.
Here $\Vec{r}=(x,y)$ denotes 2D position on the graphene
while we later use $\GVec{\rho}=(x,y,z)$ to specify the point
in three-dimensional (3D) space.
We define $\Vec{j}(\Vec{r})=(j_x,j_y)$ as the 2D electric current density
induced by the magnetic field.
Within the linear response, the Fourier-transforms of $\Vec{j}(\Vec{r})$
and $\Vec{A}(\Vec{r})$ are related by
\begin{equation}
 j_{\mu}(\Vec{q})
= \sum_\nu K_{\mu\nu}(\Vec{q})A_\nu (\Vec{q}),
\label{eq_j_a}
\end{equation}
with response function $K_{\mu\nu}$.
The gauge invariance for $\Vec{A}$ requires 
$\sum_\nu K_{\mu\nu}(\Vec{q})\, q_\nu =0$.
The continuous equation in the static system,
$\nabla\cdot\Vec{j}(\Vec{r})= 0$, imposes another constraint
$ \sum_\mu q_\mu \, K_{\mu\nu}(\Vec{q}) =0$.
To meet both requirements,
tensor $K_{\mu\nu}$ needs to be in the form,
\begin{equation}
 K_{\mu\nu}(\Vec{q}) = K(\Vec{q}) 
\left(
\delta_{\mu\nu} - \frac{q_\mu q_\nu}{q^2}
\right).
\label{eq_tensor}
\end{equation}

On the other hand, 
because $\nabla\cdot \Vec{j}(\Vec{r}) =0$, 
we can express $\Vec{j}(\Vec{r})$ as 
$j_x = c\,\partial m / \partial y$, 
$j_y = -c\,\partial m / \partial x,$
with $m(\Vec{r})$ being the local magnetic moment 
perpendicular to the layer,
and the light velocity $c$.
In the linear response, its Fourier transform is written as
\begin{equation}
 m(\Vec{q}) = \chi(\Vec{q}) B(\Vec{q}),
\label{eq_m_b}
\end{equation}
with the magnetic susceptibility $\chi(\Vec{q})$.
Equations (\ref{eq_j_a}) and (\ref{eq_m_b})
are complementary, and both response functions
$\chi(\Vec{q})$ and $K(\Vec{q})$ are related by
\begin{equation}
 \chi(\Vec{q}) = \frac{1}{cq^2} K(\Vec{q}).
\label{eq_chi_k}
\end{equation}


Graphene is composed of a honeycomb network of carbon atoms,
where a unit cell contains a pair of sublattices, denoted by $A$ and $B$.
The conduction and valence bands touch 
at the Brillouin zone corners called $K$ and $K'$ points,
where the Fermi energy lies.
The effective-mass Hamiltonian near a $K$ point
in the absence of a magnetic field
is given by 
\cite{Slonczewski_Weiss_1958a,DiVincenzo_Mele_1984a,Semenoff_1984a,Ando_2005a}
\begin{equation}
 \H_0 = \hbar v
\begin{pmatrix}
 0 & \hat{k}_x - i \hat{k}_y  \\
 \hat{k}_x + i \hat{k}_y & 0 
\end{pmatrix}
= \hbar v \hat{\Vec{k}}\cdot \GVec{\sigma},
\label{eq_H0}
\end{equation}
where $v$ is the constant velocity,
$\hat{\Vec{k}} = (\hat{k}_x, \hat{k}_y) = -i\nabla$,
and $\GVec{\sigma} = (\sigma_x,\sigma_y)$ are the Pauli matrices.
The Hamiltonian (\ref{eq_H0}) operates on a two-component wave function
$(\psi_A,\psi_B)$ which
represents the envelope functions at $A$ and $B$ sites.
The eigenstates are labeled by $(s,\Vec{k})$
with $s = +1$ and $-1$ being the conduction and valence bands, respectively, 
and $\Vec{k}$ being the wavevector.
The eigenenergy is
given by $\vare_{s\Vec{k}} = s \hbar v k$, and
the corresponding wavefunction is
$ \psi_{s\Vec{k}}(\Vec{r}) = 
e^{i\Vec{k}\cdot\Vec{r}}\, \Vec{F}_{s\Vec{k}}/\sqrt{S}$ 
with
$\Vec{F}_{s\Vec{k}} = (e^{i\theta},s)/\sqrt{2}$,
where $k$ and $\theta$ are defined by
$(k_x,k_y) = k(\cos\theta,\sin\theta)$
and $S$ is the system area.

In a magnetic field $B(\Vec{r}) = [\nabla \times \Vec{A}(\Vec{r})]_z$,
the Hamiltonian becomes $\mathcal{H} = \mathcal{H}_0 + \delta
\mathcal{H}$ 
with $\delta\mathcal{H}= (ev/c) \GVec{\sigma}\cdot \Vec{A}(\Vec{r})$.
The local current density at $\Vec{r}_0$ is calculated as
the expectation value of current-density operator
$ \hat{\Vec{j}}(\Vec{r}_0) = e v \, \GVec{\sigma} \, \delta(\Vec{r}-\Vec{r}_0)$
over the occupied states.
In the first order perturbation in $\delta{\cal H}$, we have
\begin{eqnarray}
 K_{\mu\nu}(\Vec{q})
= -\frac{g_vg_s e^2v^2}{c}\frac{1}{S}
\sum_{ss'\Vec{k}}
\frac{f(\vare_{s\Vec{k}})- f(\vare_{s'\Vec{k+\Vec{q}}})}
{\vare_{s\Vec{k}}- \vare_{s'\Vec{k+\Vec{q}}}}\nonumber\\
\times
(\Vec{F}^\dagger_{s\Vec{k}}\,\sigma_\nu\, \Vec{F}_{s'\Vec{k+\Vec{q}}})
\,(\Vec{F}^\dagger_{s'\Vec{k}+\Vec{q}}\,\sigma_\mu\,\Vec{F}_{s\Vec{k}}),
\end{eqnarray}
where $g_v=g_s=2$ are the valley ($K$,$K'$) and spin
degeneracy, respectively,
and $f(\vare) = [1+\exp \beta(\vare-\vare_F)]^{-1}$ 
with $\beta = 1/(k_B T)$
is the Fermi distribution function.

\begin{figure}
\begin{center}
 \leavevmode\includegraphics[width=\hsize]{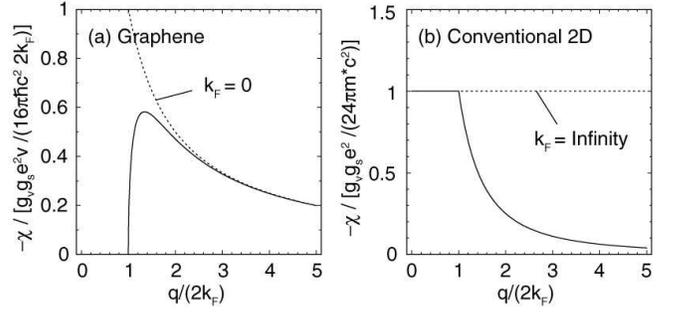}
\end{center}
\caption{
Magnetic susceptibility $\chi(\Vec{q})$
in (a) graphene
and (b) conventional 2D system.
}
\label{fig_chi}
\end{figure}

At the zero temperature, we can explicitly calculate this
to obtain
\begin{eqnarray}
 \chi(\Vec{q}\,;\vare_F) &=& -\frac{g_vg_s e^2v}{16\hbar c^2} \, 
\frac{1}{q} \,  \theta(q-2k_F) \nonumber\\
&& \hspace{-20mm}\times
\left[
1
+ \frac{2}{\pi} \frac{2k_F}{q} \sqrt{1-\Big(\frac{2k_F}{q}\Big)^2}
- \frac{2}{\pi} \sin^{-1} \frac{2k_F}{q}
\right],
\label{eq_chi}
\end{eqnarray}
where $k_F = |\vare_F|/(\hbar v)$ is the Fermi wave number
and $\theta(x)$ is 
defined by $\theta(x) = 1\,(x>0)$ and $0\,(x<0)$.
Significantly, $\chi$ vanishes 
in range $q < 2k_F$, i.e.,
no electric current is induced 
when the external field is smooth enough
compared to the Fermi wavelength.
At $\vare_F = 0$, particularly, we have
\begin{equation}
 \chi(\Vec{q}\,; \vare_F=0) = -\frac{g_vg_s e^2v}{16\hbar c^2}\frac{1}{q}.
\label{eq_chi_ef0}
\end{equation}
The susceptibility of the carbon nanotube 
to a uniform field perpendicular to the axis
has the equivalent expression of Eq.\ (\ref{eq_chi_ef0})
where $q$ is replaced by $2\pi/L$ with tube circumference $L$. 
\cite{Ajik93, Yama08}
Figure \ref{fig_chi} (a) shows a plot of $\chi(\Vec{q})$
of Eq.\ (\ref{eq_chi}).
The susceptibility suddenly starts from zero at $q = 2k_F$,
and rapidly approaches the universal curve 
(\ref{eq_chi_ef0}).
As a function of $\vare_F$ at fixed $q$,
it is nonzero only in a finite region satisfying $|\vare_F|<\hbar vq/2$,
and its integral over $\vare_F$ becomes constant 
$ -g_vg_se^2v^2/(6\pi c^2)$.
Thus, in the limit of $q \rightarrow 0$ it goes to
\begin{equation}
\chi(\Vec{q}= 0\,; \vare_F) = 
-\frac{g_vg_se^2v^2}{6\pi c^2} \delta(\vare_F).
\label{eq_chi_q0}
\end{equation}
This agrees with the susceptibility against uniform magnetic field.
\cite{McCl56,Koshino_Ando_2007b}

Let us consider an undoped graphene $(\vare_F = 0)$
under a sinusoidal field 
$B(\Vec{r}) = B_0 \cos qx$.
With the susceptibility of Eq.\ (\ref{eq_chi}),
the response current is calculated as
$ \Vec{j}(\Vec{r}) = 
- [g_vg_se^2v B_0/(16\hbar c)]\,\Vec{e}_y\,\sin qx.$
The current induces a counter magnetic field
which reduces the original field.
The $z$ component of the induced field on graphene becomes
\begin{equation}
B_{\rm ind} (\Vec{r}) = - \alpha_g B(\Vec{r}), \quad
 \alpha_g = \frac{2\pi g_vg_s e^2 v}{16\hbar c^2}.
\label{eq_alpha_g}
\end{equation}
Because the ratio is independent of $q$,
Eq.\ (\ref{eq_alpha_g}) is actually valid for
any external field $B(\Vec{r})$, i.e., 
the magnetic field on the graphene is always reduced 
by the same factor $1-\alpha_g$.
This property holds whenever $\chi(q)$ is proportional to $1/q$.
With the typical value $v \approx 1\times 10^6$ m/s,
$\alpha_g$ is estimated as $\approx 4\times 10^{-5}$,
showing that the counter field is much smaller than the original.

The argument of the magnetic field screening
can be extended in the three dimensional field distribution.
Let us suppose a situation when a certain magnetic object 
(permanent magnet or electric current) is located above 
the undoped graphene $(z>0)$,
which produces an external magnetic field $\Vec{B}(\GVec{\rho})$
in 3D space $\GVec{\rho} = (x,y,z)$.
Then, the followings can easily be concluded:
(i) On the other side of the graphene $(z<0)$,
the induced field 
becomes $-\alpha_g \Vec{B}(\GVec{\rho})$,
i.e., the external field is screened by the factor $1-\alpha_g$. 
(ii) On the same side $(z>0)$, 
the induced field is given by
$\alpha_g R_z[\Vec{B}(x,y,-z)]$,
where $R_z$ is the vector inversion with respect to $z=0$.
Namely, this is equivalent to a field of the 
mirror image of the original object reflected with respect to $z=0$,
and reduced by $\alpha_g$.


For examples, we can calculate the diamagnetic electric current
and the induced field in several specific geometries.
We first take a situation where a magnetic 
charge (monopole) $q_m$
is located at the point $\GVec{\rho}_0 = (0,0,d),\,(d>0)$ 
above the graphene plane $z=0$.
The magnetic field perpendicular to the layer on the graphene
is given by 
$B(\Vec{r}) = q_m d/(r^2+d^2)^{3/2}$
with $r=(x^2+y^2)^{1/2}$.
For $\vare_F = 0$, induced $m$ is calculated as
$m(\Vec{r}) = (\alpha_g q_m /2\pi)(r^2+d^2)^{-1/2}$,
and the corresponding current density is given by
$\Vec{j(\Vec{r})} = -c(\partial m/\partial r) \, \Vec{e}_\theta$
with $\Vec{e}_\theta$ being an azimuthal unit vector
on the $xy$ plane. 
We note that the integral of $m({\bf r})$ over the plane is
infinite, showing that it never looks like a single magnetic dipole
even if observed from far away.
Indeed, the counter field in region $z>0$ induced by $\Vec{j}$ is 
the monopole field given by $\alpha_g\,q_m$
located at $-\GVec{\rho}_0$,
as expected from the general argument above.
The monopole is thus repelled away from the graphene with a force of
$\alpha_g q_m^2/(2d)^2$.

\begin{figure}
\begin{center}
 \leavevmode\includegraphics[width=0.8\hsize]{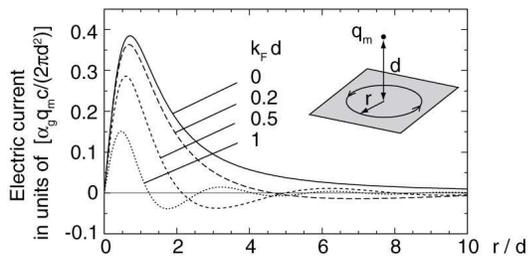}
\end{center}
\caption{
Electric current $j_\theta(r)$ on graphene
induced by a magnetic charge $q_m$ at $z=d$.
}
\label{fig_resp}
\end{figure}

For doped graphenes $\vare_F \neq 0$,
we numerically calculate the response current.
Figure \ref{fig_resp} shows $j_\theta (r)$
for several values of $k_F$.
We observe the Friedel-type oscillation with wavenumber 
$\sim 2k_F$.
The overall amplitude exponentially decays in region $k_F \agt 1/d$
because the magnetic field distribution on graphene
has a typical length scale of the order of $d$,
while $\chi(\Vec{q})$ vanishes 
in the long-wavelength region such as $q<2k_F$.


As another example, we consider
line current $I$ parallel to the graphene,
which flows along the $+y$ direction, passing through the point 
$(0,0,d)$ $(d>0)$ above graphene.
The $z$ component of the magnetic field 
on graphene is given by
$B(\Vec{r}) = -2\mu_0 I x /[c(x^2+d^2)].$
For $\vare_F = 0$, the induced current density becomes
$ \Vec{j}(\Vec{r}) =  -(\alpha_g I d /\pi)(x^2+d^2)^{-1}\, \Vec{e}_y.$
The integral of $j_y$ in $x$
exactly becomes $-\alpha_g I$, i.e.,
the external electric current 
induces an opposite current on graphene
with the amplitude reduced by $\alpha_g$.
The magnetic field induced by $\Vec{j}$ 
in the upper half space ($z>0$)
becomes equivalent to a field made by
current $-\alpha_g I$ flowing at $z=-d$, 
so that 
the original current is repelled by a force 
$\alpha_g I^2/(c^2d)$ per unit length.
When $\vare_F \neq 0$, the response current
damps for $k_F d \agt 1$,
similarly to the case of a magnetic monopole.


\begin{figure}
\begin{center}
\leavevmode\includegraphics[width=1.\hsize]{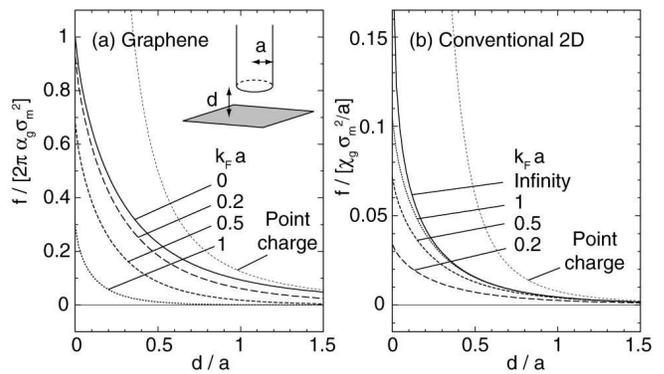}
\end{center}
\caption{Diamagnetic force per unit area
of a semi-infinite magnet cylinder with radius $a$
as a function of the distance $d$ from the tip
to (a) graphene and (b) conventional 2D system.
}
\label{fig_force}
\end{figure}


To estimate the diamagnetic force in a possible realistic situation, 
we consider a case where a semi-infinite magnet cylinder
with radius $a$,
having a flat end with surface magnetic charge density $\sigma_m$,
is placed vertically above graphene.
In real experiments, $a$ may range from
nanoscale ($\sim 10$nm) up to macroscopic length scale.
Figure \ref{fig_force} (a) shows the repulsive
force per unit area on the magnet surface,
as a function of the distance $d$ from the graphene
with several values of $k_F$.
The geometry is illustrated in the inset.
For $k_F=0$, the force is equivalent to that
made by its mirror charge $\alpha_g \sigma_m$
under graphene.
When $d\gg a$, it approaches the dotted curve given by $\propto 1/d^2$,
the force when the surface charge is replaced with
a point charge $q_m = \pi a^2 \sigma_m$. 
As $d$ goes down to the order of $a$, it deviates from
$1/d^2$ and reaches $2\pi\alpha_g \sigma_m^2$
at $d=0$, which is exactly the force between
a sheet with the magnetic charge density $\sigma_m$
and another sheet with $\alpha_g \sigma_m$,
with an infinitesimal gap.
For $\sigma_m$ which amounts to the surface flux of
1T (e.g., neodymium magnet), \cite{Coey96}
the force is estimated as $0.16$ gram force/cm$^2$,
which is surprisingly large
as a force generated by a film only one atom thick.

When $k_F$ shifts from zero, the force becomes smaller.
The tail is truncated at $d\sim 1/k_F$,
i.e., the repulsive force is lost
when the distance exceeds the order of the Fermi wavelength.
This is due to the absence of $\chi(q)$ for
long wavelength $q < 2k_F$ argued above.
Similarly, the value at $d=0$ also decays 
when $k_F$ exceeds the order of $1/a$,
because at $d=0$ the spacial distribution of $B$ on the graphene
has a typical length scale $a$.



The graphene diamagnetism is in striking contrast to 
that of the conventional 2D system.
If we apply the similar argument to Hamiltonian
${\cal H} = (\Vec{p}+e\Vec{A}/c)^2/(2m^*)$, 
the nonlocal susceptibility 
corresponding to Eq.\ (\ref{eq_chi}) yields 
\begin{equation}
 \chi(\Vec{q}\,;\vare_F) = \frac{g_v g_s e^2}{24\pi m^*c^2}
\left[
\Big(
1-\frac{4k_F^2}{q^2}
\Big)^{3/2}
\theta(q-2k_F)
-1
\right],
\label{eq_chi_conv}
\end{equation}
with $\vare_F = \hbar^2k_F^2/(2m^*)$.
The plot is shown in Fig.\ \ref{fig_chi} (b).
When $q < 2k_F$, this is constant
at $\chi_0 \equiv -g_vg_s e^2/(24\pi m^*c^2)$, 
which agrees with the usual Landau diamagnetism.
In the region satisfying $q > 2 k_F$, $\chi$ decays approximately in proportion to $1/q^2$.
In highly-doped systems such that $k_F$ is much larger than 
the typical length scale of external field $B(\Vec{r})$,
the induced magnetization just becomes $m(\Vec{r}) \approx \chi_0 B(\Vec{r})$.
For the case of a magnetic charge at $z=d$,
we have  $m(r) \propto d/(r^2+d^2)^{3/2}$, and
the integral of $m(r)$ over the plane is now finite.
The induced magnetic field at the distance $R \gg d$
is thus dipole-type decaying in proportional to $\sim 1/R^3$,
in contrast to the monopole-type field $\sim 1/R^2$
in graphene. 

Figure \ref{fig_force} (b) shows the force per unit area 
of the cylindrical magnet with radius $a$
placed above the conventional 2D system.
At large distance $d \gg a$, it rapidly decreases as $\propto 1/d^3$
like in the case of a point magnetic charge.
The peak value at $d=0$ has a typical amplitude
$\chi_0\sigma_m^2 / a$ with a factor $\sim \log k_F a$.
Apart from the factor, the ratio of the force at $d=0$
of the undoped graphene to that of the conventional 2D metal
is given by $2\pi \alpha_g a / \chi_0$,
which is roughly the ratio of the values of $\chi(q)$ at $q \sim 1/a$.
When the effective mass of GaAs $m^*\sim 0.067 m_0$
is applied to $\chi_0$,
the ratio becomes $a/$(0.01nm).
Thus, in a realistic dimension,
the diamagnetic force of the graphene is incomparably
larger than that of the conventional 2D system.
Note that, in doped graphene, the diamagnetism disappears
when $a$ becomes larger than $k_F^{-1}$.
It should be noted that $\chi(q)$ of graphene
does not approach that of conventional 2D even in the high $k_F$ limit,
because of difference between linear and quadratic dispersions.

The singular diamagnetism of graphene is influenced by temperature.
At zero doping, we expect that $\chi(q)$ deviates from 1/q
in $\hbar v q \lesssim k_B T$,
and the divergence at $q=0$ is rounded to a finite value $\propto 1/(k_B T)$.
In other words, the temperature affects diamagnetism
when the typical length scale exceeds
$2\pi \hbar v / k_B T$, which is  50 $\mu$m at $T=1$K.
We expect that the disorder potential gives a roughly similar 
effects to temperature, where the energy scale 
of level broadening works as finite $k_B T$.
\cite{Fuku07,Koshino_Ando_2007a}

The magnetism is also contributed by electron spins.
The spin susceptibility $\chi^{\rm spin}(q)$
is given by the usual density-density response function. 
\cite{Ando_2006,Brey07}
At $\vare_F = 0$, this becomes $g_v\mu_B^2 q/(16\hbar v)$ 
with the Bohr magneton $\mu_B$,
in contrast to the orbital susceptibility $\chi^{\rm orb} \propto 1/q$ 
in Eq. (\ref{eq_chi_ef0}).
The ratio $\chi^{\rm spin}/\chi^{\rm orb} \sim (q\times 0.04{\rm nm})^2$,
is quite small in realistic length scales.
We also mention that
strongly disordered graphite \cite{Esqu03} and graphene \cite{Wang09}
exhibit ferromagnetism due to spin ordering at lattice defects. \cite{Cast09}

The diamagnetism can be enhanced by stacking graphene films.
If we have randomly stacked graphenes
where the interlayer hopping is neglected,
\cite{Koto86,Hass08, Uryu_and_Ando_2005c}
the magnetic field would decay exponentially
$\propto (1-\alpha_g)^N$ with the layer number $N$.
We would have an almost perfect magnetic shield
when $N \alpha_g$ exceeds 1, which amounts to 
the thickness of 10 $\mu$m 
with the graphite interlayer spacing $0.334$ nm assumed.
We also expect that a strong repulsive force given by an external magnet
may give rise to a mechanical deformation on the graphene sheet.
The detailed study of this is left for a future work.


The authors thank R. S. Ruoff for discussions.
This work was supported in part by Grant-in-Aid for
Scientific Research on Priority Area ``Carbon Nanotube
Nanoelectronics'' and by Grant-in-Aid for Scientific Research
from Ministry of Education, Culture, Sports,
Science and Technology Japan.


\end{document}